\documentclass{revtex4}
\usepackage{amsmath, amssymb, bm, enumerate, graphicx, graphics, color,mathrsfs,hyperref}

\newcommand  {\Rbar} {{\mbox{\rm$\mbox{I}\!\mbox{R}$}}}

\newcommand{\Lie}[0]{{\cal L}\, }

\newcommand{\tl}{\theta_{(\ell)}}
\newcommand{\tn}{\theta_{(n)}}

\newcommand{\nn}{\nonumber}
\newcommand{\be}{\begin{equation}}
\newcommand{\ee}{\end{equation}}
\newcommand{\bea}{\begin{eqnarray}}
\newcommand{\eea}{\end{eqnarray}}

\newcommand{\tq}{\tilde{q}}

\newcommand{\tom}{\tilde{\omega}}

\newcommand{\vS}{\widetilde{\mbox{\boldmath$\epsilon$}}}

\newcommand{\norm}{| \mspace{-2mu} |}

\newcommand{\cV}{\mathcal{V}}

\begin{document}

\title{Horizons in the near-equilibrium regime}
\date{\today}
\author {Ivan Booth}
\email{ibooth@math.mun.ca}
\affiliation{
Departament de F\'{i}sica Fonamental, Universitat de Barcelona, \\
Marti i Franqu\`{e}s 1, E-08028 Barcelona, Spain 
\footnote{
On sabbatical leave from: \\
Department of Mathematics and Statistics, Memorial University of Newfoundland \\  
St. John's, Newfoundland and Labrador, A1C 5S7, Canada}}

\begin{abstract}
Quasi-static systems are an important concept in thermodynamics: they are dynamic but close enough to equilibrium that 
many properties of equilibrium systems still hold. Slowly evolving horizons are the corresponding concept for quasilocally defined black holes: they are
``nearly isolated'' future outer trapping horizons. This article reviews the definition and properties of these objects including both their mechanics and 
the role that they play in the fluid-gravity correspondence. It also introduces a new property: there is an event horizon candidate in close proximity 
to any slowly evolving horizon. 
\end{abstract}

\maketitle


\section{Introduction}
When trying to understand 
a physical system, 
one often starts with its equilibrium states. For example, introductions to mechanics generally 
begin with statics while the zeroth law of thermodynamics defines a system in thermal equilibrium. After the time-independent physics is 
understood, one turns to dynamics. However this generally entails a significant leap in complexity: thermodynamics is over 150 years old but 
non-equilibrium thermodynamics is still an active area of research. Thus as an intermediate step in the approach to full dynamics it is often useful 
to start with the near-equilibrium regime where many of the lessons from equilibrium physics still apply. For thermodynamics, 
this is the \emph{quasi-static} regime where a system smoothly (and generally slowly) transitions between equilibrium states.

The study of black holes follows the usual pattern. In four-dimensional asymptotically flat space-times, the equilibrium states are identified with
the Kerr-Newman family of exact solutions. Well-known theorems identify them
as the unique stationary, axisymmetric and asymptotically flat vacuum black hole solutions. As such they 
have been intensively studied since their discovery almost 50 years ago. Much of what we know about black
holes comes from a study of these equilibrium solutions and their near-equilbrium perturbations (see, for example, \cite{wald}). 

That said, as has been repeatedly emphasized elsewhere in this volume, this family of solutions is not the end of the story. Real black holes don't 
sit alone in an otherwise empty universe and so there is great interest in developing a more localized theory which characterizes black holes by their 
physical and geometric properties rather than as any particular exact solution. Hence the closely related programmes studying apparent, trapping, 
isolated and dynamical horizons (review articles include \cite{LivRev, isoreview, tomasIso, gourgoul, BHboundaries}). In this context, isolated trapping horizons are the equilibrium states of interest and dynamical trapping horizons are the corresponding 
time-dependent states.

In this chapter we will review the geometric characterization and physical properties of near-equilibirum black hole horizons. These are the 
``almost-isolated'' 
\emph{slowly evolving horizons} first defined by Booth and Fairhurst \cite{prl, bfbig}. In section \ref{characterization} we consider the geometrical 
background common to all types of horizons. With this material in hand, section \ref{GeomHor} reviews the various definitions of horizons, focussing 
on slowly evolving horizons. Section \ref{KeyProperties} discusses two of the key properties of these horizons: they obey dynamical laws of black
hole mechanics and also are always accompanied by a candidate event horizon. Section \ref{Examples} considers a couple of examples of the 
slowly evolving horizons: a simple Vaidya spacetime and some much more interesting five-dimensional black brane spacetimes that manifest the 
fluid-gravity duality. Finally section \ref{Discuss} closes this chapter with a summary of our main points.

\section{Background}
\label{characterization}

We begin with a brief review of the geometry of two- and three-dimensional surfaces in four-dimensional spacetime. Much of this material is repeated
in other chapters of this volume, however this section will serve to establish notation and perhaps present things with a slightly different perspective. 
For those who would like more details, this is essentially an abbreviated version of the extended discussion in \cite{bfbig}. 

\subsection{Intrinsic and extrinsic horizon geometry}
\label{BasicGeometry}

Let $(M,g_{ab}, \nabla_a)$ be a four-dimensional, time-orientable spacetime and $(H,q_{ab}, D_a)$ be  a three-dimensional hypersurface
which can be foliated into two-dimensional spacelike surfaces $(S_v, \tq_{ab}, d_a)$. 
The normal space to each of the $S_v$ can be spanned by a pair of future-oriented null vectors $\ell^a$ (outward-pointing) and $n^a$ 
(inward-pointing). The direction of these vectors is fixed, but their scaling isn't. One degree of freedom is usually removed by requiring that 
$\ell \cdot n = -1$ which leaves a rescaling freedom:
\be
\ell \rightarrow e^f \ell \; \; \; \mbox{and} \; \; \;  n \rightarrow e^{-f} n \, , \label{scaling}
\ee
for an arbitrary function $f$. Independent of the choice of scaling, we can write the two-metric as
\begin{equation}
\tq_{ab} = g_{ab} + \ell_a n_b + n_a \ell_b
\end{equation}
and the four-dimensional volume element $\boldsymbol{\epsilon}$ as
\begin{equation}
\boldsymbol{\epsilon} = \boldsymbol{\ell} \wedge \mathbf{n} \wedge \vS
\end{equation}
where $\boldsymbol{\ell}$ and $\mathbf{n}$ are the one-form versions of the corresponding vectors and $\vS$ is the induced
volume element on the $S_v$. 

The intrinsic geometry of the $S_v$ is determined by the induced metric $\tq_{ab}$ but the extrinsic geometry is defined by how the null 
normals vary along the surface. We have the extrinsic curvature analogues:
\begin{equation}
k^{(\ell)}_{ab} = \tq_a^c \tq_b^d \nabla_c \ell_d \; \; \mbox{and} \; \; k^{(n)}_{ab} = \tq_a^c \tq_b^d \nabla_c n_d 
\end{equation}
as well as the connection on the normal bundles:
\begin{equation}
\tom_a = - \tq_a^b n_c \nabla_b \ell^c \, . 
\end{equation}
It is useful to decompose the extrinsic curvatures into their trace and trace-free parts:
\be
k^{(\ell)}_{ab} = \frac{1}{2} \tl \tq_{ab} + \sigma^{(\ell)}_{ab} \; \; \mbox{and} \; \; k^{(n)}_{ab} = \frac{1}{2} \tn \tq_{ab} + \sigma^{(n)}_{ab} \, , 
\label{decomp}
\ee
where the traces are called \emph{expansions} and the trace-free parts are \emph{shears}.
These all depend the scaling of the null vectors and under the rescalings defined by (\ref{scaling}):
\begin{equation}
k^{(\ell)}_{ab} \rightarrow e^f k^{(\ell)}_{ab} \, , \;  \; k^{(n)}_{ab} \rightarrow e^{-f} k^{(\ell)}_{ab} \, \mbox{and} \; \; \tom_a \rightarrow \tom_a + d_a f \,. 
\end{equation}

Returning to $H$, one can further reduce the freedom allowed to the null vectors by tying their scaling to the foliation. 
Given a coordinate labelling $v$ of the foliating
two-surfaces, a unique future-oriented vector field $\mathcal{V}^a$ on $H$ satisfies the following conditions: 1) it is normal to the $S_v$, 2) it is tangent to 
$H$ and 3) it satisfies $\mathcal{L}_{\cV} v = 1$. This is the \emph{evolution vector field} for that foliation labelling: it evolves leaves of the 
foliation into each other. In coordinate language $\mathcal{V} = \partial/ \partial v$ for any parameterization in which 
 $\mathcal{V}^a$ Lie-drags the spatial-coordinates between $S_v$. Independently of whether or not such a system is constructed,
the scaling of the null vectors can be fixed by the requirement that 
\be
\mathcal{V}^a= \ell^a - C n^a \, ,
\ee
for some function $C$ which we call the \emph{expansion parameter}. Note that if $C>0$, $H$ is spacelike, 
while if $C< 0$ it is timelike and if $C=0$ it is null. We will mainly be interested in the cases where $C \geq 0$. 

Given this construction, the scaling freedom of the null vectors is restricted to the freedom to reparameterize the foliation labelling. 
For an alternative labelling $\tilde{v} = \tilde{v}(v)$ we have
\be
[d \tilde{v}]_a = \frac{1}{\alpha(v)} [d v]_a  \; \Rightarrow \; \widetilde{\mathcal{V}}^a = \alpha(v) \mathcal{V}^a
\ee
where $\alpha(v) = \frac{d v}{d\tilde{v}}$ is constant over each $S_v$. Then 
\be
\tilde{\ell}^a = \alpha  \ell^a  \, ,  \;  \tilde{n}^a = \frac{1}{\alpha} n^a \;  \; \mbox{and} \;  \tilde{C} = \alpha^2 C \, . 
\ee
Under this restricted class of rescalings with $\alpha$ constant over each individual $S_v$, $\tom_a$ is invariant. 

The intrinsic and extrinsic geometry of the full $H$ can be expressed mainly in terms of the two-surface quantities. First, the induced metric is
\be
q_{ab}  = 2 C [dv]_a [dv]_b + \tq_{ab} \, , \label{q0}
\ee
which explicitly demonstrates how the sign of $C$ determines the signature of the metric. Next, defining 
\be
\tau_a = \ell_a + C n_a 
\ee
as a future-oriented normal to the surface, the associated extrinsic curvature is
\begin{equation}
K^{(\tau)}_{ab} \equiv q_a^c q_b^d \nabla_c \tau_d = \left(2 C \kappa_{\cV} - \mathcal{L}_{\cV} C \right) [dv]_a [dv]_b + 2 C \left(\tom_a [dv]_b + [dv]_a \tom_b \right) + \left(k^{(\ell)}_{ab} + C k^{(n)}_{ab} \right) \,, \label{K0}
\end{equation}
where $\kappa_{\cV} = - \cV^a n_b \nabla_a \ell^b$. Note that we have not unit-normalized $\tau_a$ so as to allow for all values of $C$.  
If  $H$ is spacelike, the usual extrinsic curvature define in terms of the timelike unit normal $\hat{\tau}_a$ is
\be
K^{(\hat{\tau})}_{ab} = \frac{1}{\sqrt{2C}} K^{(\tau)}_{ab} \, . 
\ee

%
%

\subsection{Evolutions and deformations of two-surfaces}

Next, consider how these geometric quantities change if the surfaces are deformed. We start with the geometry of $H$ for
which the formalism will probably be more familiar as it identical to that used in the $(3+1)$-decomposition of general relativity (see, 
for example, \cite{wald, eric}). Restricting our attention to normal deformations and a spacelike $H$, potential evolutions are generated by 
vector fields of the form  $T^a = N \hat{\tau}^a$, where $N$ is a function. Then 
\begin{equation}
\delta_T q_{ab} = 2 N K^{(\hat{\tau})}_{ab} \label{dq0}
\end{equation}
while 
\begin{equation}
\delta_T K^{(\hat{\tau})}_{ab} = D_a D_b N  - N \left(R_{ab} + K^{(\hat{\tau})} K^{(\hat{\tau})}_{ab} - 2 K_a^{(\hat{\tau}) c} K^{(\hat{\tau})}_{cb}  + 
q_a^{\phantom{a} c} q_b^{\phantom{b} d} \mathcal{R}_{cd} \right) \, , \label{dK0}
\end{equation}
where as specified earlier $D_a$ is the induced covariant derivative on the surface, $R_{ab}$ is the (intrinsic three-dimensional) Ricci tensor and
$\mathcal{R}_{ab}$ is the Ricci tensor for the full
spacetime. In $(3+1)$-general relativity these are the time-evolution equations for evolutions with lapse function $N$ but vanishing shift vector. 

\begin{figure}
\includegraphics{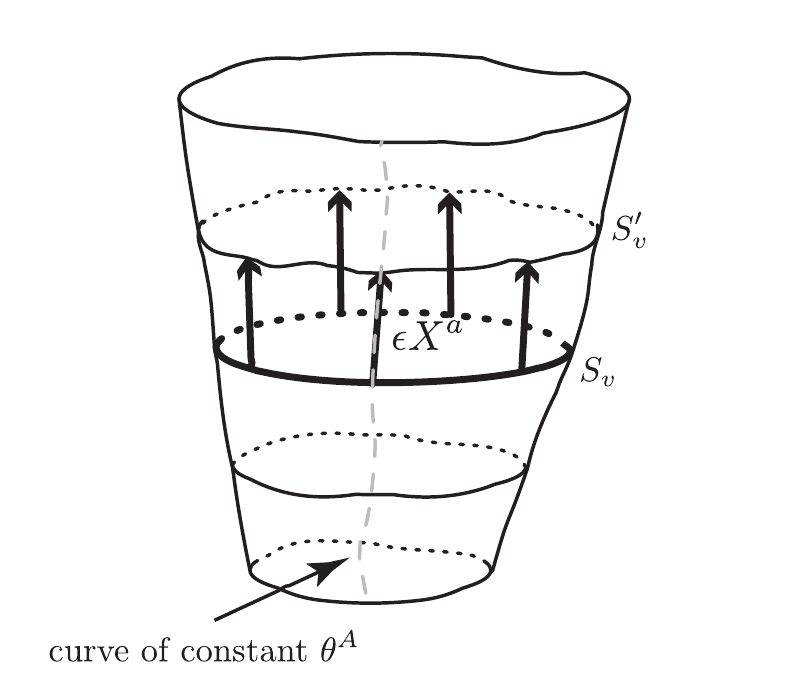}
\caption{This diagram can be interpreted in two ways. First letting $X^a= \mathcal{V}^a$ it may be thought of as a full horizon $H$ with the 
$S_v$ and $S'_v$ as foliation surfaces. Alternatively it depicts a more general deformation of an $S_v$ into $S'_v$ in which case $X^a$
is the deforming vector field. Of course the dual interpretation arises because the evolution of the horizon is just a special case of more general 
deformations.}
\label{DefFig}
\end{figure}

Similar equations can be developed for deformations of the individual two-surfaces. It is probably easiest to visualize this with the help of 
a coordinate 
system. Relative to a full coordinate system $\{ x^\alpha \}$ on some region of  $M$, we can parameterize any $S_v$ by coordinates $\theta^A$:
so that $S_v$ is embedded in $M$ by the relation $x^\alpha = \mathcal{X}^\alpha (\theta^A)$ for four functions $\mathcal{X}^\alpha$. Again we
restrict our attention to normal deformations, noting that any normal
vector field $X^\alpha (\theta)$ defined over $S_v$ can be written as 
\begin{equation}
X^\alpha = A \ell^\alpha - B n^\alpha 
\end{equation}
for some functions $A$ and $B$. Then
\begin{equation}
\mathcal{X}^\alpha (\theta^A) \rightarrow \mathcal{X}^\alpha (\theta^A) + \epsilon X^\alpha (\theta^A)
\end{equation}
defines a new surface $S'_v$ by deforming $S_v$ a coordinate distance $\epsilon$ in the direction $X^\alpha$. Further it identifies points
on the two-surfaces (essentially by Lie-dragging coordinates between surfaces) as shown in Figure \ref{DefFig}. Then, for example, the \emph{deformation}
of the two-metric $\tq_{ab}$ in the direction $X^\alpha$ is defined by
\begin{equation}
\delta_X \tq_{AB} = \lim_{\epsilon \rightarrow 0} \frac{\left. \tq_{AB} \right|_{\mathcal{X}+\epsilon X} - \left. \tq_{AB} \right|_{\mathcal{X}} }{\epsilon} \, , 
\end{equation}
where we have written the components of the metric in terms of the surface coordinates. 

As for the $(3+1)$-decomposition of general relativity, computationally this operation amounts to taking Lie derivatives with some extra conditions imposed to ensure that the geometric quantities are calculated on the correct surface and that all quantities are appropriately pulled-back into those surfaces (full details may be found in \cite{bfbig}). Then it is straightforward to show that 
\be
\delta_\ell \tq_{ab} = 2 k^{(\ell)}_{ab} \; \; \mbox{and} \; \; \delta_n \tq_{ab} = 2 k^{(n)}_{ab} 
\ee
while
\be
\delta_\ell \vS = \vS \tl \; \; \mbox{and} \; \; 
\delta_n \vS = \vS \tn \; . 
\ee
Now it is clear why the traces of the extrinsic curvatures are called expansions: they tell us how the area element changes when the surface
is evolved in the given direction. The shears are the part of the evolution that deforms the surface (but does not change its area). 

Normal deformations are linear for the first derivatives of the metric. For example:
\begin{equation}
\delta_X \vS = A \delta_\ell \vS - B \delta_n \vS = \vS(A \tl - B \tn) \label{LXep}
\end{equation}
for arbitrary functions $A$ and $B$. However, as in the three-dimensional case, things become significantly more complicated for second derivatives.
For example
\begin{eqnarray} \label{explderiv}
   \delta_X \theta_{(\ell)}&=& \phantom{-} \kappa_X \tl 
   - {d}^{\, 2} \mspace{-2mu} B  
   + 2 \tilde{\omega}^{a} {d}_{a} B \\ 
   && - B \left[ \norm \tilde{\omega} \norm^2
     -  {d}_{a} \tilde{\omega}^{a} 
     - \tilde{R} /2
     + G_{ab} \ell^{a} n^{b} - \tl \tn \right] \nn 
      - A \left[ \norm\sigma_{(\ell)}\norm^{2} 
     + G_{ab} \ell^{a} \ell^{b}
     + (1/2) \theta_{(\ell)}^{2} \right]  \, \, , 
\eea
and
\bea  \label{expnderiv}
  \delta_X \theta_{(n)}  &=& - \kappa_X \tn +
  {d}^{\, 2} \mspace{-2mu} A 
    + 2 \tilde{\omega}^{a} {d}_{a} A \\
&&   + A \left[\norm\tilde{\omega}\norm^2  + {d}_{a} \tilde{\omega}^{a} - \tilde{R}/2 
    + G_{ab} n^{a} \ell^{b} - \tl \tn \right] \nn
  + B \left[ \norm\sigma^{(n)}\norm^{2} 
    + G_{ab} n^{a} n^{b} 
    + (1/2)\theta_{(n)}^{2} \right] \,  
      \, .
\eea
where $\kappa_X = - X^a n_b \nabla_a \ell^b$,  $\tilde{R}$ is the two-dimensional Ricci scalar for $S_v$, $\norm \tom \norm^2 = \tom^a \tom_a$,  
$\norm \sigma_{(\ell)} \norm^2 = \sigma_{(\ell)}^{ab} \sigma^{(\ell)}_{ ab} $ and $\norm \sigma_{(n)} \norm^2 = \sigma_{(n)}^{ab} \sigma^{(n)}_{ ab}$.
For the connection one-form:
\begin{eqnarray}  
\delta_{X} \tilde{\omega}_{a} 
     &=&  -  ( A \tl - B \tn) \tom_a  + {d}_{a} \kappa_{\mathcal{V}} 
    -  {d}_{b} \left( A \sigma_{(\ell) \; a}^{\; \; b}  
      + B \sigma_{(n) \; a}^{\; \; b} \right) -  \tilde{q}_{a}^{\; b} G_{bc} \tau^{c}  \label{ang_mom_expression1}
        \\
    &&\qquad  + \frac{1}{2} {d}_{a} \left( A \theta_{(\ell)} + B \theta_{(n)} \right) - \theta_{(\ell)} {d}_{a} A
    - \theta_{(n)} {d}_{a} B  \nonumber  \, , 
\end{eqnarray}
where, as before, $\tau^c = A \ell^c + B n^c$ is normal to $X^c$. All of these second derivatives depend on how 
$A$ and $B$ vary over the $S_v$.  

Probably the most intuitive application of these deformations is in understanding the time-evolution of a horizon. Then $X^a$ is identified with 
the $\mathcal{V}^a$ evolution vector field and, for example,
\begin{eqnarray} 
   \delta_{\cV} \theta_{(\ell)}&=& \phantom{-} \kappa_{\cV} \tl 
   - {d}^{\, 2} \mspace{-2mu} C  
   + 2 \tilde{\omega}^{a} {d}_{a} C  \label{explVderiv} \\ 
   && - C \left[ \norm \tilde{\omega} \norm^2
     -  {d}_{a} \tilde{\omega}^{a} 
     - \tilde{R} /2
     + G_{ab} \ell^{a} n^{b} - \tl \tn \right] \nn 
      -  \left[ \norm\sigma_{(\ell)}\norm^{2} 
     + G_{ab} \ell^{a} \ell^{b}
     + (1/2) \theta_{(\ell)}^{2} \right] \, \, . 
\eea
In particular for a null surface $\mathcal{\cV}^a = \ell^a$ and we recover the Raychaudhuri equation
\be
\delta_\ell \theta_{(\ell)} = \kappa_\ell \tl 
      - \left[ \norm\sigma_{(\ell)}\norm^{2} 
     + G_{ab} \ell^{a} \ell^{b}
     + (1/2) \theta_{(\ell)}^{2} \right]  \, \, , \label{ray}
\ee
which is key to many of the calculations involving isolated \cite{tomasIso} or event \cite{wald,eric} horizons. 

\section{Geometric horizons}
\label{GeomHor}
We are now ready to define slowly evolving horizons. In this section we will use geometric arguments, 
leaving physical properties for the next section. We begin with a review of the definitions of isolated and dynamical trapping horizons.
%
%

\subsection{General quasilocal horizons}
\label{GQH}

In a four-dimensional spacetime $(M,g_{ab}, \nabla_a)$ a \emph{future outer trapping horizon (or FOTH)} is a three-surface $H$ that is foliated by 
spacelike two-surfaces $(S_v, \tq_{ab}, d_a)$ such that on each surface: i) $\tl = 0$, ii) $\tn < 0$ and iii) there is a positive function $\beta$ such that
$\delta_{\beta n} \tl < 0$ \cite{hayward}. 
These conditions are intended to (locally) mimic those used to define apparent horizons \cite{hawkellis}: each slice of a FOTH is marginally 
outer trapped ($\tl = 0$) but the other two conditions guarantee that it is possible to deform the $S_v$ inwards so that they become fully trapped. 

That the inward 
expansion be negative is straightforward to check and independent of the scaling of the null vectors, however the last condition is not quite so simple. 
From the general deformation equation for $\tl$ we see that with $\tl = 0$:
\be
\delta_{\beta n} \tl =  d^2 \beta - 2 \tom^a d_a \beta -  ( \tilde{R} /2 + \norm \tilde{\omega} \norm^2
     -  {d}_{a} \tilde{\omega}^{a} 
     + G_{ab} \ell^{a} n^{b} ) \, \beta  \label{dntl} \, .
\ee
Now, for a given scaling of the null vectors, deciding whether condition iii) holds will amount to a study of the possible behaviours of this second
order elliptical differential operator. Even for fairly simple cases, a non-trivial $\beta$ can be required. 
For  example, rapidly rotating Kerr solutions and the standard scaling of the null vectors obtained from Boyer-Lindquist co-ordinates has 
$\delta_n \tl  \nless 0$ (appendix C of \cite{bfbig}). However, the horizons certainly are FOTHs and this can be demonstrated with a suitable choice of 
$\beta$. Similarly for general FOTHs away from spherical symmetry, it will often be the case that the scaling determined by fixing 
$\mathcal{V}^a = \ell^a - C n^a$, is not a scaling for which $\delta_n \tl < 0$. Again a non-trivial $\beta$ is required.

Since $\tl = 0$ on a FOTH we much also have $\Lie_\cV \tl = 0$. Then applying a maximum principle argument to (\ref{explderiv}) 
along with the fact that $\delta_{\beta n} \tl < 0$ (for some $\beta$) one can show that the null energy condition implies that $C\geq 0$ which in turn 
means that 
\be
\Lie_\cV \vS = - C \tn \vS \geq 0 \, , \label{SecondLaw}
\ee
 since $\tn < 0$. Thus,
there is a second law of FOTH mechanics: their area is non-decreasing. 

Next, we summarize isolated horizons (see \cite{tomasIso} in this volume for a more information or \cite{isoMech} for the full details). 
A three-dimensional submanifold in that same spacetime is a \emph{non-expanding horizon} if: 
i) it is null and topologically $S \times \Rbar$ for some closed two-manifold $S$, ii) $\tl = 0$ and 
iii)  $-T^{ab} \ell_b$ is future directed and causal. As usual $\ell^a$ is a future oriented, outward-pointing normal; since the horizon is 
null, in this case no foliation is needed for its construction. That said a foliation is certainly no hindrance to a three-surface being a non-expanding
horizon and any null FOTH satisfying the null energy condition will certainly be a non-expanding horizon. 

Non-expanding horizons are the simplest objects in the isolated horizon family. Any non-expanding 
horizon can be turned into a \emph{weakly isolated} horizon if the scaling of the null vectors is chosen so that
$\Lie_\ell \omega_a = 0$ where
\bea
\omega_a \equiv - n_b \nabla_{\underleftarrow{a}} \ell^b = - \kappa_{\ell} n_a + \tom_a \, , 
\eea
and the arrow indicates a pull-back into the cotangent bundle of the non-expanding horizon. With 
this scaling, zeroth and first laws of isolated horizon mechanics may be established. Finally, if
there exists a scaling of the null vectors for which the entire extrinsic geometry of the horizon is 
invariant in time then the weakly isolated horizon is an \emph{isolated horizon}. Thus in that case not only
do $\Lie_\ell \kappa_{\ell}$ and $\Lie_\ell \tom_a$ vanish, but so do $\Lie_\ell \tn$ and $\Lie_\ell \sigma^{(n)}_{ab}$.
All Killing horizons are isolated in this sense.


Our final object is a \emph{dynamical horizon}\cite{ak}.
A three-dimensional sub-manifold of a spacetime $(M,g_{ab})$ is a dynamical horizon 
if: i) it is spacelike and ii) can be foliated by spacelike two-surfaces such that the null normals 
to those surfaces satisfy $\tl = 0$ and $\tn < 0$. Perhaps the key property of dynamical horizons
is that they always expand: the spacelike assumption means that $C > 0$ and so by the strictly greater-than version of (\ref{SecondLaw}), dynamical 
horizons always expand. 
Not surprisingly, if the null energy condition holds and $\delta_\ell \tl \neq 0$ at least somewhere on each $S_v$, then 
a FOTH is a dynamical horizon.
Another important property of dynamical horizons is that their foliation into $S_v$ surfaces with $\tl = 0$ is \emph{unique} \cite{Ashtekar:2005ez}.
This is
very convenient as we do not need to worry about whether or not geometric properties are foliation dependent: they 
are, but since the foliation is unique this is fine!

Note that this foliation rigidity contrasts with the corresponding situation for isolated horizons where foliations can be freely deformed. 
However, the situation is very different if we consider the rigidity of the full three-dimensional $H$. 
Marginally outer trapped surfaces in an isolated horizon can only be deformed along the horizon itself. By contrast for a dynamical FOTH, 
the $\tl = 0$ surfaces can only be deformed \emph{out of} $H$ (this is essentially equivalent to how the exact location of a dynamical apparent horizon 
depends on the foliation of the full spacetime). For an extended discussion of this point see \cite{bfbig}.



We will take FOTHs as our basic objects and classify them by hybridizing the  naming systems as was done in \cite{bfbig}. 
Thus a FOTH that also satisfies one of these other sets of properties will be referred to as 
non-expanding, (weakly) isolated, or dynamical as appropriate. There are also more exotic forms of FOTHs associated with apparent horizon 
``jumps'' (see for example \cite{bendov, mttpaper}), inner horizons or white holes \cite{hayward}. However, they can be left aside for the purposes
of this article. The horizons that we will be interested in will always be dynamical or weakly isolated FOTHs. 

\subsection{Near-equilibrium quasilocal horizons}

Intuitively, a near-equilibrium black hole boundary should be ``almost'' isolated. Thus, since isolated horizons are non-expanding and null, 
a near-equilbrium
horizon should expand slowly and be almost null. The complication here is quantifying ``expand slowly'' and ``almost null'': expanding dynamical horizons 
are, by their very nature, spacelike so it is not immediately clear what ``slowly'' means and further it is certainly not obvious how a spacelike surface can
be ``almost null''. 

In analogy with isolated horizons, we break the definition up into two parts: \emph{slowly expanding horizons} which focus on restrictions to the intrinsic 
geometry and \emph{slowly evolving horizons} which add constraints to the extrinsic geometry. 

\subsubsection{Slowly expanding horizons}


\vspace{0.3cm}
\noindent
{\bf Definition:} Let $\triangle H$ be a section of a future outer trapping horizon foliated by two-surfaces 
$S_v$ so that
$\triangle H = \{ \cup_v S_v : v_1 \leq v \leq v_2\}$. Further let $\cV^a$ be an evolution vector field 
that generates the
foliation so that $\Lie_{\cV} v = \alpha(v)$ for some positive function $v$, and scale the null vectors 
so that 
$\cV^a = \ell^a - C n^a$. Finally let $R_H$ be the characteristic length scale for the problem.
Then $\triangle H$ is a \emph{slowly expanding horizon} if the dominant 
energy condition holds and
\begin{enumerate}
\item $\displaystyle |\tilde{R}|, \tom_a \tom^a, | d_a \tom^a |$ and $T_{ab} \ell^a n^b \lesssim 1/
R_H^2$.
\item  Two-surface derivatives of horizon fields are at most of the same magnitude as the (maximum)
of the original fields. For example, $\displaystyle \norm d_a C\norm \lesssim C_{max}/R_H$, 
where $C_{max}$ is largest absolute value attained by $C$ on $S_v$.
\item $\epsilon \ll 1$ where 
$\epsilon^2/R_H^2 =  \mbox{Maximum} \left[ C \left( \norm \sigma^{(n)} \norm ^2 + T_{ab} n^a n^b + \tn^2/2 \right) 
\right]$. 
\end{enumerate}
If the $\alpha(v)$ is chosen so that $C \approx \epsilon^2$ then this scaling of the null normals is 
said to be compatible with the \emph{evolution parameter} $\epsilon$. \\

This definition requires some discussion and we analyze it point-by-point. 
Starting with the preamble, we specify a section $\Delta H$ rather than a full horizon so that the definition will
cover sections of horizons that may be slowly expanding only for finite periods of time. 
For standard four-dimensional black holes the characteristic length scale $R_H$ will 
be set by the areal radius $R_H =  \sqrt{A/(4 \pi)}$.

Next, following from the discussion back in section \ref{BasicGeometry},
 choosing $\mathcal{L}_{\cV} v = \alpha (v)$ for some  positive function $\alpha(v)$ ensures that the flow generated by $\cV^a$ evolves 
leaves of the horizon into each other and so can be thought of as a coordinate vector field 
compatible with the foliation. The free function is included so that the scaling is not tied to a particular choice of foliation labelling; 
including it is equivalent to allowing for relabellings of the $S_v$.

Turning to the main clauses, the notation $X \lesssim Y$ means that $X \leq k_o Y$ for some constant $k_o$ of order one. Then our first 
condition that  
\bea
| \tilde{R} | \, , \,  \tom^a \tom_a \, , \, |d_a \tom^a|  \, \mbox{and} \, |G_{ab} \ell^a n^b| \lesssim \frac{1}
{R_H^2} \, , 
\label{Rt2dtTln}
\eea
is essentially a restriction on the allowed geometries of the horizons. The first part ensures that the curvature is not too extreme, the second two 
assumptions place some restriction on the extrinsic curvature while the last bounds the Einstein tensor (or equivalently stress-energy). In particular
it was shown in \cite{bfbig} that these restrictions hold for all members of the Kerr family of horizons. An immediate consequence of these
assumptions is that by (\ref{dntl})
\bea
|\delta_n \tl |  \lesssim \frac{1}{R_H^2} \, , \label{dntlBnd}
\eea
is similarly bounded. Note the use of the absolute value sign. Even though we must have $\delta_{\beta n} \tl < 0$ for $\beta$ scaling, in general this will 
not be true for $\beta=1$. 

The second condition restricts the size of the derivatives of the horizon fields
so that they are at most commensurate with their maximum values. For tensor quantities such as $\sigma^{(n)}_{ab}$ the magnitudes are taken to be
judged relative to an orthornormal frame. For example
\begin{equation}
\left\| d_a \sigma^{(n)}_{bc} \right\| \lesssim  \frac{1}{R_H}  \left\| \sigma^{(n)}_{bc} \right\|
\end{equation}
indicates that this inequality should hold for all components relative to any orthonormal frame on $S_v$. 


We can then consider the last condition. After the set-up of the earlier conditions, this is the one that captures the essence of what it means 
to be slowly evolving. Beginning with the evolution of the area-form $\vS$ on $S_v$, by (\ref{LXep}) we have
\bea
\Lie_{\cV} \vS = - C \tn \vS \, , 
\eea
which is certainly dependent on the scaling of the null vectors. However this dependence may 
be easily isolated by rewriting
\bea
\Lie_{\cV} \vS = - C \tn \vS = || \cV ||  
\left(- \sqrt{\frac{C}{2}} \tn \vS \right) \, ,
\eea
so that the effects of the rescaling freedom are restricted to pre-factor $||\cV|| = \sqrt{2C}$. The term in
parentheses (which is required to be small by the third condition) then provides an invariant measure of 
the rate of expansion. Among other properties, it vanishes if the horizon is non-expanding while
on a dynamical horizon section it is equal to the rate expansion of $\vS$
with respect to the unit-normalized version of the evolution vector
field \footnote{One might be concerned as to whether this quantity 
vanishes in the limit $C \rightarrow 0$ (in which case $\widehat{\cV}^a$ is no longer defined). It does. 
For an explicit demonstration see the discussion in \cite{PhysQuant}.}.
The examples of section  \ref{Examples} provide intuitive support for the identification of 
small $\epsilon$ with a slow expansion. 


This notion of a slow area change is the key part of the entire definition. 
The rest of the third condition forces the surface to be
``almost-null" with a slowly changing intrinsic metric. To simplify
notation we now adapt a evolution-parameter-compatible scaling of the null vectors so that 
\begin{equation}
C \approx \epsilon^2 
\end{equation}
which also ensures that any transition to equilibrium and so a fully isolated horizon will be smooth.
With this scaling the first two conditions along with (\ref{explderiv}) imply that 
\begin{equation}
| \delta_\ell \tl | = \sigma^{(\ell)}_{ab} \sigma_{(\ell)}^{ab} + T_{ab} \ell^a  \ell^b \lesssim \frac{\epsilon^2}{R_H^2} \, . 
\end{equation}
and so the rate of expansion becomes 
\bea
\Lie_{\cV} \vS = \underbrace{- C \tn \vS}_{O(\epsilon^2)} \, .
\eea
while the ``time''-derivative of the full metric is 
\bea
\Lie_{\cV} \tq_{ab} = 
   \underbrace{2 \sigma^{(\ell)}_{ab}}_{O(\epsilon)} 
  + \underbrace{\left( C \tn \tq_{ab} - 2 C\sigma^{(n)}_{ab}
\right)}_
  {O(\epsilon^2)} \, . \label{LcVex}
\eea
The leading terms are the same as they would be for a truly null surface. Further the leading terms of the energy flux across the horizon 
\bea
\cV^a T_a^b \tau_b \approx  \underbrace{T_{ab} \ell^a \ell^b}_{O(\epsilon^2)} -  \underbrace{C^2 T_{ab} n^a n^b}_{O(\epsilon^4)}
\, , \label{MatFlux}
\eea
(where $\tau_a = \ell_a + C n_a$ is the usual timelike normal) also match those of a truly null surface.  

We refer the reader to \cite{bfbig} for further discussion of these and other consequences of the definition.

%
%
%
%
%
%

\subsubsection{Slowly evolving horizons}
For slowly expanding horizons we imposed conditions to ensure that $\triangle H$ was almost null and that the two-geometry changed only
slowly. We now move on to slowly evolving horizons where additional conditions are imposed to also restrict the evolution of much of the rest of the 
horizon geometry. This can be motivated by the analogous development of the isolated horizon formalism from non-expanding horizons
(restricting intrinsic geometry) to fully isolated horizons (also restricting extrinsic geometry). 

The two-metric $\tq_{ab}$, extrinsic curvatures ($k^{(\ell)}_{ab}$, $k^{(n)}_{ab}$) and normal connection $\tom_a$ fully specify the geometry of the $S_v$. The rest of the geometry of $H$ is then fixed by 
$\kappa_{\cV}$ and $C$. From the previous section, 
for a slowly expanding horizon with an adapted evolution parameter, we have a good notion of a 
``time''-derivative. Thus, for these remaining quantities we can directly bound their rates of change with respect to $\mathcal{V}^a$, demanding that
in each case their rate of change be of a lower order than the original quantity. 

\smallskip
\noindent
{\bf Definition:} Let $\triangle H$ be a slowly expanding section of a
FOTH with a compatible scaling of the null normals. Then it is said to
be a \emph{slowly evolving horizon} (SEH) if in addition

\begin{enumerate}
\item $\displaystyle \norm\Lie_{\cV} \tom_a \norm$ ,  $|\Lie_{\cV}
\kappa_{\cV}|$ and  $|\Lie_{\cV} \tn| \lesssim \epsilon/R_H^2$ and 
\item $|\Lie_{\cV} C| \lesssim {\epsilon^3}/{R_H}$. 
\end{enumerate}

Note that the inclusion the $\mathcal{L}_{\cV} C$ is new in this article as compared to earlier definitions of slowly evolving horizons. It is 
not required for the first law however, as we shall see, it is required to prove that there is always an event horizon candidate in close proximity to 
any SEH.


\section{Key Properties}
\label{KeyProperties}

In this section we review two of the key properties of slowly evolving horizons. 

\subsection{Laws of Mechanics}

Slowly evolving horizons are intended to be the black hole analogues of quasi-static systems in thermodynamics. As such we would expect them to 
(approximately) obey the laws of black hole mechanics (a nice discussion of the equilibrium forms of these laws can be found in the isolated
horizon chapter of this book which is also online at \cite{tomasIso}). 

Starting with the zeroth law, we expect that surface gravity will be approximately constant on each surface $S_v$. This is the case. Applying
the various defining conditions to the equation for the variation of $\tom_a$ (\ref{ang_mom_expression1}), it is immediate that
\be
\norm d_a \kappa_{\cV} \norm \lesssim \frac{\epsilon}{R_H^2} 
\ee
(the analogous equation for null surfaces is used to show that the surface gravity is constant for a (weakly) isolated horizon).  
Given that the surface gravity has already been required to been slowly changing up the horizon, it follows that over a foliation
parameter range on $\triangle H$ that is small relative to $1/\epsilon$, 
\begin{equation} \label{kap0}
  \kappa_{\cV} = \kappa_o + {O}(\epsilon)
\end{equation}
for some constant $\kappa_o$. Note however that, as for temperature in a standard quasi-static process, the surface gravity can accumulate larger
changes over sufficiently long periods of time. 
 
We have already seen that FOTHs that satisfy the null energy condition are non-decreasing in area. SEHs inherit this property. This is one
form of the second law, however given that SEHs represent quasi-static black holes one might also expect them to dynamically satisfy the 
second law in its Clausius form.
For reversible thermodynamic processes, entropy is defined by 
\be
\delta S =  \frac{\delta Q}{T} \, . \label{clausius}
\ee
Switching to black hole mechanics we expect the analogous law to link area increase, energy flux and surface gravity.

Slowly evolving horizons
do obey a law of this form \cite{prl,bfbig}. First linearly combining (\ref{explderiv}) and (\ref{expnderiv}) one can show that
\bea
\kappa_{\cV} \theta_{\cV} = \delta_\cV \tl + B \delta_\cV \tn 
      + d_a (C d^a C - 2C \tom^a) \label{kapT} 
   + \sigma^{(\tau)}_{ab}  \sigma_{(\cV)}^{ab}
   + G_{ab} \cV^a \tau^b 
   +  \frac{1}{2} \theta_{(\cV)} \theta_{(\tau)} \, , \nn
\eea
where the shears associated with $\cV^a$ and $\tau^a$ are defined in the obvious way. Integrating over $S_v$ eliminates the total derivative 
term and then applying slowly expanding/evolving conditions (including the zeroth law) we find that to order $O(\epsilon^2)$:
\begin{equation}\label{FirstLaw}
  \frac{\dot{a}}{4 \pi G} \approx 
 \frac{2}{\kappa_o} \int_{S_v} \vS \left\{\frac{\norm\sigma^{(\ell)}\norm^2}{8 \pi G} 
  + T_{ab} \ell^a \ell^b \right\} \, . 
\end{equation}
Inside the integral on the right-hand side, the terms can respectively be identified as fluxes of gravitational radiation and stress-energy and, as 
would be expected, they are equivalent to the terms that one finds for truly null surfaces (see for example the discussions in \cite{hawkhartle, eric}). 

Note, though we have called this the Clausius form of the second law, it is more traditional to refer to it as the first law of black hole mechanics\cite{BHMech}. For stationary black holes and isolated horizons this first law takes the form
\be
\delta M = \frac{1}{8 \pi G} \kappa \delta A + \Omega \delta J + \Phi \delta Q \, ,
\ee
where the $M$ is the black hole mass (taken as the internal energy), $J$ is angular momentum and $Q$ is the charge while $\Omega$ and $\Phi$ 
are their associated potentials. However, it has been pointed out \cite{GourJara} that while there are energy flow terms in  (\ref{FirstLaw}),
there is no explicit splitting of those terms into work terms versus those involving changes in the internal energy of the system. The notion of internal 
energy is key to the thermodynamic interpretations of the first law and as such it probably makes more sense to view the energy flow terms as more
analogous to the heat flow in (\ref{clausius}). Hence we refer to it as a form of the second law.

%
%
%
%
%
%
%
%
%
%
%
\subsection{Spacetime near an SEH}

In this section we will demonstrate that (at least to leading order) there is an actual null surface close to each slowly evolving FOTH. 
This null surface is a candidate event horizon and is, in fact, the actual event horizon if the FOTH remains an SEH forever. Further, despite the
teleological definition of event horizons the behaviour of this event horizon candidate is locally determined and its evolution closely mirrors that of the 
SEH. Full details of this result will appear in \cite{nearSEH}.

%
%

\subsubsection{Event horizons}

\begin{figure}
\scalebox {1.1}{\includegraphics{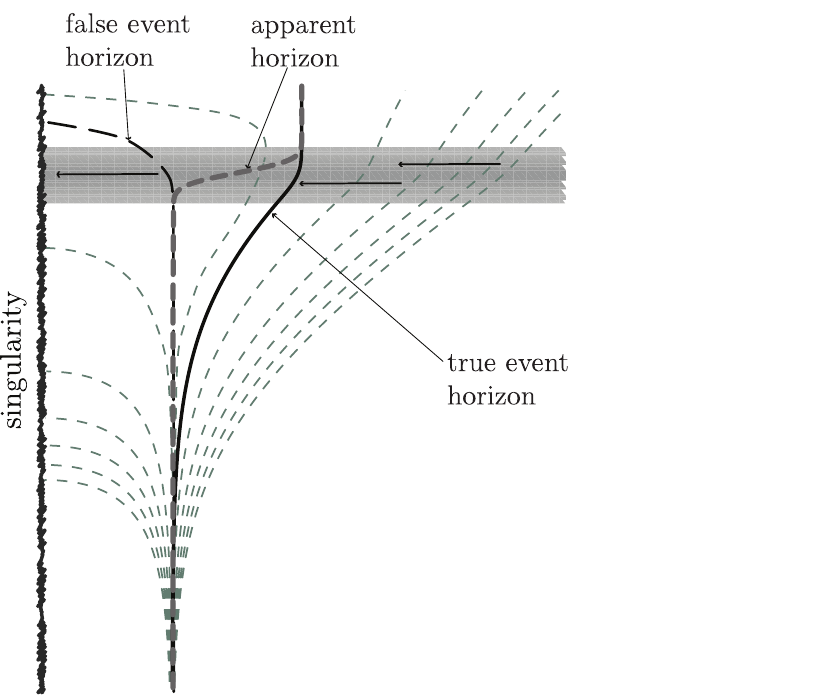}}\caption{A schematic that plots both the quasilocal and event horizons for a typical Vaidya spacetime
in which a shell of dust (the shaded gray region) falls into a pre-existing black hole. In this figure, horizontal location measures the areal radius of 
the associated spherical shell while the direction of increasing time is roughly vertical outside the event horizon but 
tipping horizontal-and-to-the-left inside. On both sides, inward-moving null geodesics are horizontal while ``outward-pointing'' null geodesics 
are represented by gray dashed lines. The false event horizon is a null surface that would appear to be the event horizon if one is
unaware that the matter shell will fall into the black hole in the future and so capture it.}
\label{EHFig}
\end{figure}
An event horizon is the boundary of a \emph{causal black hole}: a region of spacetime from which no causal signal can escape. 
Such a surface is necessarily null and, for outside observers, it is the boundary between the unobservable events inside the black hole and 
those outside that can be seen. This definition is teleological: one determines
 the extent (or existence) of a black hole by tracing all causal paths ``until the end of time'' and then retroactively identifying any black hole region.
The most obvious manifestation of this type of definition is that, as shown in Figure \ref{EHFig}, infalling matter curtails rather than causes
the expansion of an event horizon.

The reason for this behaviour is mathematically easy to understand. As a null surface, the event horizon is necessarily ruled by null geodesics. 
Thus, in the language of earlier sections, it is a three-surface with $C=0$ and so the Raychaudhuri equation holds:
\begin{align}
\Lie_{\cV} \tl =& \kappa_{\ell} \tl  - \left( {\tl^2}/{2}   + \norm \sigma_{(\ell})\norm^2 + G_{ab} \ell^a \ell^b \right)   \label{explderiv2} \, . 
\end{align}
Now, given the Einstein equations and null energy condition, if we temporarily adopt an affine scaling of the null vectors ($\kappa_\ell = 0$), it follows that 
\bea
\Lie_\ell \tl \leq - \frac{1}{2} \tl ^2  \label{explderiv3} \, . 
\eea
Thus the expansion naturally decreases with time. There is nothing profound about this statement. Consider for example a
sphere of light expanding outwards in flat space so that $R(t)= R_o + ct$. Then 
\be
\tl = \frac{2 \dot{R}}{R} = \frac{2c}{R(t)}
\ee 
and so the rate of expansion decreases as time (and therefore) $R$ increases: the rate of expansion scales with the (inverse) radius of the horizon. 
Returning to general relativity a non-zero flux of gravitational energy or positive-energy density matter will cause the rate to decrease even more. 
Again this is to be expected: the gravitational influence of more mass inside the shell should decrease the rate of expansion. 

The apparently counterintuitive nature of event horizon evolution is then seen to be not so mysterious. Event horizons are families of null geodesics 
evolving exactly as null geodesics should. The particular set of geodesics has been chosen based on future boundary conditions 
(not escaping to infinity and not falling into the singularity) but at any given moment they evolve causally just like any other null congruence. 

The real mystery is then: why is it that numerical \cite{NumEx} and perturbative calculations \cite{PertEx} often show event horizons evolving
in the apparently intuitive way, expanding in response to infalling radiation and matter? Rewriting the Raychaudhuri equation as
\begin{equation}
\kappa_\ell \tl - \Lie_\ell \tl = \norm \sigma_{\ell} \norm^2 + G_{ab} \ell^a \ell^b + \frac{1}{2} \tl ^2 \label{Raych}
\end{equation}
it is straightforward to see that just a couple of slowly-evolving-type conditions will be sufficient to enforce a Clausius-type second law (a.k.a the first law
of black hole mechanics). For a small $\tl$ one would have $\tl^2 \ll \tl$ while one might also expect the rate of change of the expansion to be smaller than
the expansion itself $\Lie_\ell \tl \ll \tl$ (as in a standard derivative-expansion like those considered in the fluid-gravity correspondence \cite{Mukund}). 

We now consider how to express this intuition in geometrically invariant way.
First note that on right-hand side of (\ref{Raych}) the $\tl^2$ term can be neglected if 
\be
 \frac{1}{2} \tl ^2 \lesssim \epsilon \left( \norm \sigma_{\ell} \norm^2 + G_{ab} \ell^a \ell^b \right) \, ,
 \ee
for some $\epsilon \ll 1$. 
Both sides of this inequality change in the same way with respect to scalings of the null vectors so this condition is geometrically meaningful and 
can be used to determine the scale of ``slowness''. Next, if we can scale the null vectors so that on the left-hand side
\be
\Lie_\ell \tl \lesssim \left( \frac{\epsilon}{R_H} \right) \tl \, , 
\ee
and $\kappa_\ell$ is of order $1/R_H$, then the Clausius-type law is recovered for this null surface:
\be
\kappa_\ell \tl \approx  \norm \sigma_{\ell} \norm^2 + G_{ab} \ell^a \ell^b \, .
\ee
Thus, in this limit fluxes will appear to drive the expansion of these null surfaces in the intuitive way.
Further discussion of these slowly evolving null surfaces can be found in \cite{vaidya}.

\subsubsection{Event horizon candidate near an SEH}
Intuitively one might expect a truly null surface to lie close to any slowly evolving horizon. In fact, this is the case and that null surface is also slowly 
evolving in the sense discussed above. To first order this is easily seen. 
As noted earlier, the extrinsic curvature of a hypersurface is the rate of change of the induced metric under normal deformations. 
Then, the induced metric on a three-surface close to a slowly evolving horizon is 
\be
q^{(\Delta \tau)}_{ab} \approx q^{(0)}_{ab} + (\Delta s) \left[  \Lie_{\hat{\tau}} q_{ab}  \right]^{(0)}=  q^{(0)}_{ab} + 2 (\Delta s) K^{(0)}_{ab} \, ,
\ee
where $\hat{\tau}$ is the future-oriented timeline normal to $\triangle H$ and $\Delta s$ measures proper time along the corresponding timelike
geodesics. Substituting in (\ref{q0}) and (\ref{K0}) this becomes 
\bea
q^{(\Delta \tau)}_{ab} \approx \left(2 C +  (\Delta s) \sqrt{2C} \bar{\kappa}_{\cV} \right) [dv]_a [dv]_b
+ \sqrt{2C} (\Delta s) \left(\tom_a [dv]_b + [dv]_a \tom_b \right) + \left(\tq_{ab} +  \frac{\Delta s}{\sqrt{2C}} \left(k^{(\ell)}_{ab} + 2 C k^{(n)}_{ab}  \right)
\right) \, , \label{FindEH}
\eea
where 
\be 
\bar{\kappa}_{\cV} \equiv \kappa_{\cV} - (\Lie_{\cV} C)/\sqrt{2C} \approx \kappa_{\cV}
\ee
(by assumption the derivative is of lower order than $\kappa_{\cV}$ and so may be neglected). 

Thus, the surface a proper time 
\be 
\Delta s \approx -\sqrt{2C}/\kappa_{\cV} \label{EHLoc}
\ee 
into the past of the SEH will be null to leading order. It is not hard to 
demonstrate that this surface is also slowly evolving in the sense considered in the last section and so evolves in a local way which closely
mirrors that of the SEH \cite{nearSEH}. If $C$ remains small and ultimately asymptotes to zero then this null surface also 
asymptotes to the SEH and truly is the event horizon.

It is important to keep in mind that even in this case, the event horizon is still teleologically defined. We identify it based
on the assumption that it will remain slowly evolving for all eternity. If, at some point in the future the FOTH is no longer slowly 
evolving, then the construction breaks down and this set of null geodesics will dive into the singularity as did the false horizon 
in Figure \ref{EHFig}. 

\section{Examples and Applications}
\label{Examples}

In order to build some intuition about these horizons, we now turn to a couple of concrete examples. The first is very simple, an SEH in a 
Vaidya spacetime, and it mainly serves to help to establish the range of astrophysical processes that we might expect to be slowly evolving. 
The second example is an SEH in a five-dimensonal black brane spacetime. This is technically much more complicated and also has the 
very interesting property that it is dual to a four-dimensional conformal fluid through the fluid-gravity duality. As such we can use it transfer lessons
from slowly evolving horizons to fluid dynamics and vice versa.  

\subsection{Vaidya black holes}
\label{BillsEx}
We begin with a spherically symmetric example in four-dimensions: slowly evolving Vaidya black holes. This is taken
from \cite{billspaper} which discusses both this example and some other, slightly more realistic, ones. As in that paper 
we are primarily interested in using the example to build physical intuition and so take take one solar mass as our standard unit:
\be
M_\odot = 1.9 \times 10^{30} \, \mbox{kg} \, .
\ee
Hence, with $G=6.7 \times 10^{-11} \, \mbox{m}^3 \, \mbox{kg}^{-1} \, \mbox{s}^{-2}$ and 
$c = 3.0 \times 10^8\,  \mbox{m} \,  \mbox{s}^{-1}$, 
distances and times that come out of our equations will be measured in 
units of
\bea
R_{\odot} &=& GM_\odot/c^2 \approx 1.4 \times 10^3 \,  \mbox{m}  \, \, \, \mbox{and} \\
T_\odot & = & R_\odot/c \approx 4.7 \times 10^{-6} \, \mbox{s} \,  
\eea
respectively. Relative to everyday experience, our unit of mass is huge while our unit of time is very small. As such we might expect to find that 
some fairly dramatic events lie in the slowly evolving regime. We will see that this is the case. 



\subsubsection{The spacetime}
\label{vaidya}

The Vaidya solution \cite{eric} is the simplest example of a dynamical black hole spacetime.
It models the collapse of null dust and is described by the metric
\bea
ds^2 &=& - \left(1- \frac{2m(v)}{r} \right) dv^2 + 2 dv dr + r^2 d \Omega^2 \, , 
\eea
where $v$ is an advanced time coordinate, $m(v)$ measures the mass of the black hole on a hypersurface of constant  $v$ and the infalling null dust has stress-energy tensor  
\bea
T_{ab} &=& \frac{dm/dv}{4 \pi r^2} [dv]_a [dv]_b \, .
\eea
Figure \ref{EHFig} shows one such a spacetime in which a shell of null dust accretes onto a pre-existing black hole. 

To calculate the various quantities of interest we fix a scaling of the null vectors to surfaces of constant $r$ and $v$:
\bea
\ell^a = \left[1, 1 - \frac{2m(v)}{r} , 0 , 0  \right] \, \, \,  \mbox{and} \, \, \,  n^a = [0, -1, 0, 0] \, .
\eea
Then straightforward calculations show that for these spherically symmetric two-surfaces:
\bea
\tl = \frac{1}{r} \left( 1 - \frac{2m(v)}{r} \right) \, \, \, \mbox{and} \, \, \, \tn = - \frac{2}{r} \, 
\eea
while
\bea
\kappa_{\ell} = - n_b \ell^a \nabla_a \ell^b =\frac{m}{r^2}  \; \mbox{and} \; G_{ab} \ell^a \ell^b = \frac{2 \dot{m}}{r^2} \, . 
\eea
Overdots indicate derivatives with respect to $v$. 
Thanks to both the symmetry and the special nature of null dust, most other geometrical quantities (including the shear and all other components
of the stress-energy tensor) vanish. 

\subsubsection{The slowly evolving horizon}

Now, there is clearly a FOTH located at $r = 2m(v)$ with characteristic length $R_H = 2 m$. For this surface 
\bea
C =   2 \dot{m}
\eea
so 
\bea
\epsilon^2 \equiv \frac{1}{2} C \tn^2 (R_H)^2 = 4 \dot{m} 
\eea
and it is straightforward to see that if $\dot{m} \ll 1$, then this is slowly expanding horizon. Next, 
\bea
\kappa_{\cV} = \frac{1}{4m}  \; \; \mbox{and} \; \; \mathcal{L}_\cV C = 4 \ddot{m} \, . 
\eea
and so, if $\ddot{m} \ll \dot {m}$ the horizon is also slowly evolving.

Before checking implications, we restrict to a specific example to get a feeling for the kind of situations under which Vaidya is slowly evolving. 
Consider a piecewise linear mass function : 
\bea
m(v) = \left\{
\begin{array}{ll}
1 & v \leq 0 \\ 
1+ \alpha v & 0 \leq v \leq v_o\\
1+ \alpha v_o & v>v_o
\end{array} \right.  \, .
\eea
That is, we irradiate an initially solar mass black hole with null dust for a finite period of time. 
 
To interpret this physically,  consider a fleet of observers who use rockets to hold themselves some distance from the horizon
at constant areal radius $r_o$. These observers will see a total mass of $\alpha v_o$ fall past them during time 
\be
T =  \int_0^{v_o} \sqrt{1 - \frac{2m(v)}{r}} dv <  \alpha v_o \, , 
\ee 
with the inequality coming close to saturation for very large $r_o$. Of course, these observers can't actually see the horizon but since it is spacelike
there aren't any suitable observers who actually live on the horizon. 

Adding in some numbers, consider $\alpha = 1/40000$ and $v_o \approx 2.1 \times 10^5 M_\odot$ (one second in standard
units). Then, our observers would see $\alpha v_o \approx 5.3$ solar masses of material sweep past 
them in less then a second. However, at the horizon a quick calculation shows that 
$\epsilon^2  = 4 \alpha = 10^{-4}$. Thus, even in this very dramatic situation the horizon would 
be slowly evolving to order $\epsilon \approx 0.01$. 

There are many possible objections to this example, from the fact that null dust doesn't exist to the assumption of spherical symmetry. However,
it is in line with our expectations based on the units $M_\odot$ and $T_\odot$ and broadly comparable results can be found in \cite{billspaper} 
which includes timelike-dust examples and also Schwarzschild as perturbed by quite strong incoming gravitational waves. Details vary, but the 
general message is that black holes probably evolve slowly in almost all astrophysical situations, with one notable exception being black hole mergers. 

\subsubsection{Mechanics}

For spherically symmetric Vaidya, the zeroth law holds trivially (and exactly) while the FOTH at $r=2m(v)$ trivially is clearly increasing in area
for $\dot{m} > 0$. Thus the only part of the mechanics that needs checking is the Clausius form of the second law (\ref{FirstLaw}). 
From the preceding section 
\be 
\frac{\kappa_o \dot{a}}{8 \pi G} \approx \frac{dm}{dv} 
\ee
while the right-hand side is also 
\be 
 \int_{S_v} \vS \left\{\frac{\norm\sigma^{(\ell)}\norm^2}{8 \pi G} + T_{ab} \ell^a \ell^b \right\} \approx \frac{dm}{dv}\, . \label{VaidyaFirst}
\ee
In fact as was discussed in \cite{billspaper} the Vaidya spacetime is, in many ways, not the best spacetime to demonstrate that this law holds:
thanks to the symmetries and special nature of null dust, (\ref{VaidyaFirst}) actually holds exactly whether or not the 
horizon is slowly evolving. This however is an exceptional circumstance
and for all other examples (including Tolman-Bondi) the slowly-evolving condition is necessary.

\subsubsection{The event horizon candidate}
\label{vaidyaEH}

For the Vaidya spacetime one can directly search for null surfaces just outside the event horizon by trying to solve
\begin{equation}
\frac{dr}{dv} = \frac{1}{2} \left(1 - \frac{2m}{r} \right)  
\end{equation}
for a solution of the form $r = 2m (1 + \rho)$, with $\rho \ll 1$. If one assumes a hierarchy of derivatives 
(so that $m \gg m \dot{m} \gg m^2 \ddot{m}$), one can iteratively solve this equation order-by-order to any desired
accuracy.  In particular to second order one finds that \cite{Mukund}:
\begin{equation}
r_{null}  \approx  2m +   8 m \dot{m} + \left( 64 m \dot{m}^2 + 32 m^2 \ddot{m} \right) \, . 
\end{equation}
If this coordinate-dependent distance is converted into a proper-time measurement \cite{vaidya}, it matches the first order prediction 
of Eq.~(\ref{FindEH}).

\subsection{Black branes in the fluid-gravity correspondence}
\label{FluidEx}

A much more interesting example of near-equilibrium horizons can be found in the context of the fluid-gravity correspondence. 
This is not the place to go into a full discussion of this version of the AdS-CFT correspondence, but very
briefly it states that a near-equilbrium asymptotically AdS black hole in $N$-dimensions is dual to a near-perfect conformal fluid in
$(N-1)$-dimensions. From the CFT perspective, these near-equilbrium black holes correspond to the long-wavelength regime
where the field theory reduces to fluid mechanics. That said, from a purely gravitational perspective, there is no need to reference
the full correspondence. A direct examination of the long-wavelength perturbations of AdS black branes demonstrates that
the gravitational equations of motion governing those perturbations are equivalent to the Navier-Stokes equations governing a 
near-perfect conformal fluid \cite{Mukund}. 

Thermodynamically the temperature of this fluid should correspond to that of the black brane while the entropy flow should map onto the 
expansion of its horizon, and indeed this is the case to leading order. However, at higher order things become more interesting
and several ambiguities arise in the exact meaning of the phrase ``the entropy flow should map onto the 
expansion of its horizon''. 

In this section we briefly review the role of slowly evolving horizons in the fluid-gravity correspondence. Further details can be found in
\cite{fg1,fg2,fg3}.  Up until now we have restricted our attention to four-dimensional spacetimes with spatially compact black holes. 
Switching to five-dimensions for this example, the formalism that we have developed remains essentially unchanged. The only changes are that 
we relax the 
compactness condition (with the scale now being set by the surface gravity rather than the areal radius) and have to replace some factors of 
 $1/2$ with $1/3$ in equations such as (\ref{decomp}),  (\ref{explderiv})-(\ref{explVderiv}) and (\ref{ray}).

\subsubsection{The spacetime}

Starting with a five-dimensional AdS black-brane metric, the fluid gravity correspondence may be derived in the following way \cite{Mukund}. 
This solution is static and on inducing a compatible time-foliation, each instant may be thought of as a stack of three-dimensional intrinsically flat planes that 
stretch from the spacetime singularity to infinity with a similarly planar event horizon in between. In Eddington-Finkelstein-like form the metric is 
\be
ds^2 = - r^2 \left(1  - \frac{1}{(rb_o)^4} \right) dv^2 + 2 dv dr + r^2 (dx^2 + dy^2 + dz^2) \, ,
\ee
where for computational convenience the cosmological constant has been scaled to take the value $-6$ and the event horizon at $r=1/b_o$ has 
temperature
\be
T = \frac{1}{\pi b_o} \, . 
\ee
An alternate form of this metric arises if one performs a Lorentz boost within the planes so that we switch to a coordinate frame which moves with 
four-velocity $u_o^\alpha$ relative to the static system:
\be
d s^{2} = - r^{2} \left( 1 - \frac{1}{\left( r b_o
  \right)^4} \right) u^o_{\mu} u^o_{\nu} d x^{\mu} d x^{\nu} +  2 u^o_{\mu} d x^{\mu} d r  + r^{2} \left(
\eta_{\mu \nu} + u^o_{\mu} u^o_{\nu} \right) d x^{\mu} d x^{\nu}. \label{boostedbrane}
\ee
In this case the greek indices run over $\{v, x, y, z\}$ and $\eta_{\mu \nu}$ is the standard four-dimensional Minkowski metric. 

The fluid-gravity correspondence comes from considering long-wavelength perturbations of this boosted black brane metric: one replaces the 
constant $b_o$, $u_o^\alpha$ and $\eta_{\mu \nu}$ with $b$, $u^\alpha$ and $g_{\mu \nu}$ which agree with their unperturbed counterparts at 
leading order but beyond that have a coordinate dependence. In the planar directions the scale of variation of these perturbations must be much 
longer than the length scale set by $b$ (they are long-wavelength perturbations). Thus, for example,
\be
\norm d_\alpha u^\beta \norm \ll \left\|  \frac{u^\beta}{b} \right\| \, . 
\ee
Further, one assumes that this pattern continues for the entire hierarchy of derivatives in the planes. Continuing with the $u^\beta$ example:
\be
\norm u^\delta \norm \gg b \norm d_\alpha u^\delta  \norm \gg b^2 \norm d_\alpha d_\beta  u^\delta \norm \gg b^3 \norm d_\alpha d_\beta d_\gamma  u^\delta \norm \dots
\ee
It is then possible to separate and solve the Einstein equations for these perturbations order-by-order. As might be expected, though straightforward in 
principle, these computations rapidly become very intricate\cite{Mukund,bhatta2}.

For those interested in horizons these approximate solutions are of interest in their own right. At second order in this perturbation theory we can 
track the evolution of both event and apparent horizons and they are distinct from the unperturbed results. By contrast for Schwarzschild spacetimes
current state-of-the-art perturbation theory cannot distinguish these horizons and places them both at $r \approx 2m$; corrections to the position are 
at a higher order than can be tracked \cite{ericOld,ericNew1,ericNew2}. 
However, before considering those horizons in more detail we complete our discussion of the fluid-gravity correspondence as  
it turns out that the conformal fluid interpretation eases the calculations on the gravity side. 

The perturbed solution is entirely described by the quantities $\{b, u_\mu, g_{\mu \nu} \}$ with all tensor-indices being in the cotangent space 
to the surfaces of constant $r$. Each of these surfaces has an intrinsic geometry defined by $g_{\mu \nu}$ (it is weakly perturbed from Minkowski) 
and on each such surface one can construct a four-dimensional stress-energy tensor:
\be
T^{\mu \nu} = \epsilon u^{\mu} u^{\nu}  + p h_{\mu \nu}  \ ,
\ee
where
\be
h_{\mu \nu} = g_{\mu \nu} + u_{\mu} u_{\nu}
\ee 
is the three-metric transverse to the $u_\mu$ and the pressure and energy density are defined by $b$ and obey the equation of state:
\be
\epsilon = 3 \, p =  \frac{3}{16 \pi G b^4}  \, .\label{equationofstate}
\ee
This may be regarded as a formal definition (which will be justified by later developments), however it may also be derived systematically as a 
Brown-York boundary stress-energy tensor\cite{BrownYork}. In any case, for the unperturbed metric this is the stress-energy of a conformally 
invariant perfect fluid. However on implementing the perturbations it is only near-perfect with the corrections depending on gradients of 
$\{b, u^\mu, g_{\mu \nu} \}$ (and corresponding to non-perfect-fluid corrections such as viscosity). 

Implementing the Einstein equations turns out to be equivalent to requiring conservation of the stress-energy tensor on each of these 
surfaces:
\be
\mathcal{D}_\mu T^{\mu \nu} = 0 \, ,
\ee
where $\mathcal{D}_\mu$ is the conformally invariant covariant derivative compatible with $g_{\mu \nu}$ (see the original sources or \cite{fg3} for more
details).
This can be observed directly from computation with the components, but it may also be seen as a consequence of the fact that the momentum constraint 
must hold on each surface (again see \cite{BrownYork}).  

The dual fluid is generally taken to live on the surface at $r \rightarrow \infty$.
The conformal invariance can be used to rescale away diverging quantities in this limit and so guarantee that the theory remains well-defined.

\subsubsection{The slowly evolving horizon}

For the Vaidya example, finding a FOTH was straightforward; guided by the spherical symmetry, we simply examined the surfaces of constant $r$
(and ignored all other candidates). Here things are a bit more complicated. The geometric symmetry of the problem is broken and so one cannot use it 
to select and search for a preferred FOTH. However, there is still the conformal symmetry and this turns out to be sufficient to select a preferred FOTH
(to second order). Again we will be brief, leaving the reader to consult \cite{Mukund,bhatta2, Loganayagam:2008is, fg3} for more details. 

The conformal symmetry of the fluid means that solutions are invariant under rescalings of the form
\be
g_{\mu \nu} \rightarrow e^{- 2\phi} g_{\mu \nu}, \quad u^{\mu} \rightarrow
e^{\phi} u^{\mu}, \ \quad b \rightarrow e^{-\phi} b \quad \mathrm{and} \quad r
\rightarrow e^{\phi} r \, . 
\ee
Physically meaningful quantities should also respect this symmetry and we take this to include the location and evolution vector field of the FOTH. 
This is a strong restriction as there are only a limited number of conformally invariant quantities available at each order of the gradient expansion. 
For example, at first order, there is only the Weyl-invariant shear of $u^\mu$:
\begin{equation}
\sigma^{fluid}_{\mu \nu} = \mathcal{D}_{(\mu} u_{\nu)} \, ,
\end{equation}
where the parentheses indicate the usual symmetrization.  At second order in gradients there are ten quantities 
(three scalars, two vectors, and five transverse traceless tensors). We require that corrections to both the location of the horizon and its
evolution vector field $\mathcal{V}^\mu$ be constructible out of these quantities. 

After some fairly extensive algebra one can show that to second order there is a unique conformally invariant FOTH at 
\begin{equation}
r_H = \frac{1}{b} + \frac{1}{b} \left(h^{AH}_1 S_1 - \frac{3}{8} S_2 - \frac{1}{24} S_3 \right)  \label{FOTH}
\end{equation}
where $h^{AH}_1$ is a non-trivial numerical factor, 
while the conformally invariant factors are built out of the conformally invariant shear and vorticity of $u^\mu$ as well as the similarly conformally 
invariant curvature scalars: $S_1 = b^2 \sigma_{\mu \nu} \sigma^{\mu \nu}$, $S_2  = b^2 \omega_{\mu \nu} \omega^{\mu \nu}$ and
$S_3  = b^2 \mathcal{R}$. 
Note that the corrections are all at second order in gradients. At first order the horizon is still at $r=1/b$. 
This FOTH is easily seen to be slowly evolving with characteristic length scale $R_H = 1/b$. 

The exact implications of this uniqueness are unclear. It has been conjectured \cite{fg3} that the uniqueness might extend to higher order corrections 
and possibly even imply that in this case the conformal invariance selects a unique FOTH just as in the Vaidya case the spherical symmetry selected
a preferred horizon. If true this would be a fascinating result, however it is also possible that the uniqueness to second order is simply a consequence 
of the fact that the horizon is slowly evolving. The resolution of this question will be left to future investigations.

\subsubsection{Mechanics: horizon and fluid}

On the gravity side it is easy to check that both the (approximate) zeroth law holds ($\kappa_{\cV} \approx 1/b$) as does the Clausius form of the second
law. However for these solutions there is an additional interest as we can compare the mechanics of the near-equilibrium horizon to the
thermodynamics of the similarly near-equilbrium fluid. By the fluid-gravity correspondence we would expect them to match. 

Before we can do this, we first need to understand the thermodynamics of the fluid. It is standard to identify the fluid temperature with that of the horizon
so that to first order 
\be
T_{fluid} \approx \frac{1}{b} \, . 
\ee
Then the equation of state (\ref{equationofstate}) defines energy density and pressure as functions of temperature. 

The next step is to identify an entropy current $J^\mu$: the divergence of such a current $\mathcal{D}_\mu J^\mu$ determines the rate of entropy 
production in the fluid. For equilibrium flows this takes the form 
\be
J_{eq}^\mu = s u^\mu
\ee
where $s$ is the (equilibrium) entropy density. However, away from 
equilibrium the correct form of the entropy current is not so clear and it is usually argued that 
one should allow for correction terms \cite{Mukund, Romat, fg1,fg2}. In our situation we
must consider all conformally invariant causal flows with non-negative divergence. As for the horizons, requiring that the flow be 
conformally invariant significantly restricts the freedom of construction, but in this case is does not completely remove it: there are allowed 
second order corrections to $J_{eq}^\mu$ and these correspond to different values of entropy production.

The rate of expansion of the slowly evolving FOTH defined by (\ref{FOTH}) corresponds to one of these values of entropy production. Given that FOTHs are 
non-unique (recall that they are not rigid and may be deformed), it is very tempting to then conjecture that the other $J^\mu$ correspond to other (nearby) FOTHs. However, this cannot be the case: 
even though the conformal symmetry does not select a unique entropy current, it did select a preferred slowly evolving horizon. Thus any 
gravitational interpretation of this uncertainty must be sought elsewhere. 

\subsubsection{The event horizon candidate}

Our FOTH is slowly evolving, and so we would expect to find a (similarly slowly evolving) event horizon candidate in close proximity. 
This is the case \cite{Mukund} and so this candidate provides another potential gravitational dual to the entropy current. Indeed it is straightforward to 
check that the rate of expansion of the event horizon candidate is matched by the divergence of another member of the family of allowed entropy currents. 

\subsubsection{Implications and complications}

From the horizon perspective, these black brane spacetimes with their (measurably) dynamical horizons are nice examples in their own right providing 
testing grounds for ideas about slowly evolving horizons. The fluid gravity duality adds another level of interest to these examples. In the 
first place we can see that dual to the near-equilibrium horizons are the similarly near-equilibrium fluids and that the mechanics of the horizon is 
similarly dual to a thermodynamics of the fluid. 

Note the phrase ``a thermodynamics'' rather than ``the thermodynamics''. The entropy flow in the fluid is not uniquely defined
and so therefore neither is its thermodynamics. Both the FOTH and event-horizon candidates have their own (closely-related but distinct) thermodynamic
duals. However, these are just two of the potential thermodynamics and so it is natural to consider duals to the
other entropy flows. There are at least two possible ways to account for this freedom in a geometric way:
\begin{enumerate}
\item The conformal fluid lives on the AdS boundary of the spacetime and should be dual to the entire spacetime, not just the horizon. It can be argued that
the brane dynamics dominate the spacetime, however even in that case one must still define a map between the horizon and the boundary. Given the 
coordinate system, it is computationally easiest to just match up points with matching $x^\mu$ coordinates.
However this mapping is not geometrically preferred and there are other acceptable ways to set it up.  It turns out that the freedom in constructing 
the mapping closely matches that in 
defining the entropy flow\cite{Mukund}. So, the first possible geometric explanation for the entropy flow freedom is that it is a manifestation of the 
horizon-boundary mapping uncertainty. Either the FOTH or the event horizon may be viewed as dual to each of the entropy flows. 
\item The second possibility is to take the freedom on the fluid side as signalling that a similar freedom should exist in defining 
entropy and entropy flows in the bulk. References \cite{fg1,fg2} suggest one way of doing this in which the SEH and  
event horizon candidate are taken to be just two examples of a family of horizon-like structures any one of which can be associated with 
a mechanics (and each of which is dual to a fluid flow). These surfaces can have any signature but each signals the existence of a nearby 
casual boundary and is (at least approximately) null and slowly expanding with $\tl \approx 0$. In any approach to equilibrium 
they all are required to have the same isolated horizon limit. In particular the timelike ones are similar to those surfaces considered in the 
membrane paradigm \cite{membrane}. 
\end{enumerate}
Of course these two possibilities are not mutually exclusive. If the conformal fluid reflects the true physics of the system, then the first resolution means that
the horizon-candidates proposed by the second resolution are indistinguishable since both have the same range of freedom in their definition. Thus each 
candidate would furnish an equally acceptable dynamical black hole mechanics. These would differ in detail but remain closely related and in particular
all reduce to the same equilibrium limit. 

A definitive resolution of these issues will have to wait for future work. However it may be worth pointing out that 
uncertainties of this type would not be too surprising.  While there are several well-defined measure of total energy in general relativity \cite{wald},
its localization is famously ill-defined and the study of potential quasi-local definitions of energy remains an area of active research \cite{szabados}. 
Thermodynamics and black hole mechanics are intimately tied to energy flows and so if there are uncertainties in amounts of energy transferred
it would be consistent if there were similar uncertainties in entropy.  


\section{Summary}
\label{Discuss}

The first law of black hole mechanics as formulated for exact solutions or isolated horizons is a phase space relation: it considers variations through 
the phase space of equilibrium states rather than a dynamical evolution between those states. The original motivation for the investigation of slowly 
evolving horizons was to better understand the near-equilibrium regime for dynamical trapping horizons and so reinterpret this first law in a dynamical 
way \cite{prl}. We have seen that this reformulation is possible and the resulting dynamical version arises as a particular limit of the equations 
governing the geometric deformation of the surfaces. 

For null surfaces the deformation equations reduce to the Raychaudhuri equation and so also govern the evolution of event horizons, though the 
identification of any particular null surface as the event horizon remains a teleological procedure. There is also a slowly evolving limit for null surfaces
and we saw that there is a surface of this type in close proximity to each slowly evolving horizon. If the FOTH remains near-equilibrium at all 
times in the future, that null surface will be the event horizon. Its physical properties and geometry are very similar, though not identical, to those 
of the slowly evolving horizon. 

A wide variety of astrophysical processes are probably in the near-equilibrium regime. One can quickly see this by noting that the characteristic
time scale for black hole would be the horizon-crossing time. For a solar mass black hole this is on the order of $10^{-6}$ seconds. Thus we 
would expect anything that happens on a time scale much greater than this to be slowly evolving. Particularly dramatic examples were 
demonstrated in the Vaidya spacetime. 

Finally we considered the interesting case of five-dimensional black branes and the fluid-gravity duality. In that case the perturbatively constructed
spacetime gives rise to a slowly evolving horizon in the bulk and a corresponding near-equilibrium fluid on the boundary, with the mechanics of the 
horizon closely matching the thermodynamics of the fluid. However, the thermodynamics is not uniquely defined and can be adjusted to match both 
the SEH and its attendant event horizon candidate. It can be matched with the evolution of other nearly null, nearly isolated ``horizons''. This 
behaviour is reminiscent of the classical membrane paradigm which noted that much black hole physics can be modelled by considering the
interaction of near-horizon timelike surfaces with their  environment. 

\section*{Acknowledgements} This work was supported by the Natural Sciences and Engineering Research Council of Canada.

\end{document}